\documentstyle[preprint,aps]{revtex}
\begin{document}
\title{Reading the three--dimensional structure of a protein from its amino
acid sequence}
\author{R. A. Broglia$^{1,2,3}$ and G. Tiana$^3$}
\address{$^1$ Dipartimento di Fisica, Universita' di Milano, via Celoria 16,
20133 Milano, Italy}
\address{$^2$ INFN, sezione di Milano, via Celoria 16, 20133 Milano, Italy}
\address{$^3$ The Niels Bohr Institute, Bledgamsvej 17, 2100
Copenhagen, Denmark}
\date{\today}
\maketitle

\begin{abstract}
While all the information required for the folding of a protein is contained
in its amino acid sequence \cite{anfinsen}, one has not yet learnt how to extract
this information so as to predict the detailed, biological active,
three--dimensional structure of a protein whose sequence is known
\cite{creighton,branden,fersht}.
This situation is not particularly satisfactory, in keeping with the fact
that while linear sequencing of the amino acids specifying a protein is
relatively simple to carry out, the determination of the
folded--native--conformation can only be done by an elaborate X--ray
diffraction analysis performed on crystals of the protein or, if the protein
is very small, by nuclear magnetic resonance techniques. Using insight
obtained from lattice model simulations of the folding of small proteins
(fewer than 100 residues), in particular of
the fact that this phenomenon is essentially controlled by conserved
contacts \cite{mirny} among strongly interacting amino acids, which also stabilize
local elementary structures \cite{baldwin} formed early in the folding process
and leading to the (post--critical) folding core when they assemble together
\cite{aggreg}, we have worked out a successful strategy for reading the
three--dimensional structure of a notional protein from its amino acid
sequence.
\end{abstract}

\bigskip
\bigskip

An answer to the question of how to relate the protein primary structure to
its full three--dimensional structure has long been awaited to come from a
proper understanding of the mechanisms which are at the basis of the
phenomenon of protein folding \cite{creighton2}. 
The fulfillment of such expectation is likely
to revolution the therapeutic drug industry, let alone the chemistry of
enzymatic processes. It may also be within reach, in any case, to the
extent that simple lattice models of protein folding do capture
some of the essential properties of real proteins. 

Although Anfinsen and collaborators had conclusively shown \cite{anfinsen} that the
1D--structure of a protein determines its 3D--structure, researchers had
long been stymied in their efforts to predict the latter from the knowledge
of the former, because of Levinthal's paradox \cite{levinthal}: the number of all
possible conformations of a polypeptide chain is too large to be sampled
exhaustively. Nevertheless, proteins do fold into unique native states in
seconds.

A major breakthrough in the study of protein folding was made by inventing a
simple (although not oversimplified) model of protein folding, 
the so called inverse folding
model \cite{sh_prl}, in which the quest of the relation existing between the 1D--
and the 3D-- structure of a notional protein is formulated in such a way
that Levinthal's paradox is circumvented. In essence, this model turns the
problem of protein folding inside out. First, a native conformation is chosen
and then, the notional protein designed by minimizing the energy of the
system with respect to the amino acid sequence (for fixed composition).
Using a 20--letter three dimensional lattice model for heteropolymers
\cite{sh_prl,GSW,SSK1,PNAS,GOLDSTEIN,LENGTH,BRYN_REVIEW,COSB,PANDE}, 
and contact energies obtained from a statistical analysis of real
proteins \cite{mj}, one finds that good--folder--sequences are characterized by
a large gap $\delta=E_n-E_c$ (compared to the standard deviation $\sigma$ of
the contact energies) between the energy $E_n$ of the designed sequence in
the native conformation and the lowest  energy (threshold $E_c$) of the
compact conformations structurally dissimilar to the native conformation. In
other words, Monte Carlo (MC) simulations testify to the fact that
designed sequences displaying a large normalized gap ($\xi=\delta/\sigma\gg
1$, quantity closely related to the z--score \cite{bowie}) in the native conformation
fold on short call \cite{note1}.
The success of the inverse folding model is connected with the fact that
good folders share a (small) number of conserved contacts (which eventually
form the folding nucleus of the protein), contacts which act among the most
strongly interacting and best conserved amino acids
(so called "hot" and "warm" sites in ref. \cite{Tiana_98}).

We now know \cite{aggreg,prlnuovo} 
that this result is tantamount to saying that foldability is
controlled by the presence of few, local elementary structures, stabilized by
the conserved contacts and formed very early in the folding process, leading
to the (post--critical) folding core (i.e., the minimum set of native
contacts needed to ensure foldability \cite{sh_nature,nucleus})
 when they assemble together.
From this vantage point of view it is easy to see how to solve the
"inverse--inverse folding problem", that is how to predict the 3D--structure
of a protein, from its 1D--structure (for details we refer to Methods):
1) starting from a notional sequence and making use of the contact energies
used in its design find the local elementary structures,  2)
determine the possible folding cores by allowing the local elementary
structures to interact among them, 3) relax the position of the remaining
amino acids and determine the corresponding energy. The (single) compact
structure which displays an energy smaller than $E_c$ is the native
conformation. This in keeping with the fact that all sequences which in the
compactation process display local elementary structures and thus a folding
core will fold to a unique conformation, provided the associated total energy
lies below $E_c$ \cite{prlnuovo}.
We have applied this strategy to representative members of essentially all
the classes of lattice designed sequences available in the literature:
27mers \cite{27_1,27_2,48_1}, 36mers 
\cite{aggreg,Tiana_98,desig_prl,klimov2,socci,s2,thirumalai}
48mers \cite{sh_nature} and 80mers \cite{sh_prl}. In all
cases the predicted native structure coincides with the compact structure
used to design the protein by carrying on it simulated annealing in
sequence space.

An example of the results 
obtained by applying the strategy described above to a designed
36mer is shown in Fig.1. The results displayed in Figs. 1(b)
and 1(c) testify to the fact that the designed sequence shown in Fig.
1(a) is a good candidate to code for a notional protein. In fact, the energy
associated with the contacts stabilizing the three predicted local elementary
structures (yellow dashed lines, Fig. 1(b) ) is $-2.66$ in the units we are
using ($RT_{room}=0.6$ kcal/mol), the interaction energy among these
structures in the (post--critical) folding core
(red dashed lines, Fig. 1(c) ) being equal to $-5.15$, the 
total energy of this core
thus being $-7.81$. Relaxing the amino acids not contained in the core, we
find for the sequence shown in Fig. 1(a) that the system can lower its
energy below $E_c$ (to a value -16.5, where $E_c$,
calculated making use of the Random Energy Model \cite{derrida} is $-14.1$) for the
conformation shown in Fig. 1(d), which is thus predicted to be the native
conformation of the sequence shown in Fig. 1(a). In fact, this
conformation coincides with the one used in the literature to design the 
sequence under discussion. This sequence, known as $S_{36}$,
folds into the native conformation shown
in Fig. 1(d) with an (average) first passage time of $0.71 \cdot 10^6$ MC steps,
following the hierarchy of steps described above (formation of elementary
structures in $\approx 10^2$ MC steps, formation of the (post--critical) folding
core in $\approx 0.7 \cdot 10^6$ MC steps and folding in $\approx 0.71\cdot 10^6$ MC steps). 
These steps can be also observed in
Fig. 1(e), where the time dependence of the native contacts 
for a particular run used to calculate the average folding time of the protein
is shown as a
function of the number of steps of the Monte Carlo simulation.   

As documented by the results displayed in 
Fig. 2, even the 36mer sequence (cf. Fig. 2(a) ) designed
onto a conformation which displays
as few local contacts as possible \cite{nonlocal}, 
surrenders its 3D structure (cf. Fig.
2(d) ) to the local elementary structure--strategy discussed above. 
In this case, only one of the local elementary structures is closed, i.e.
contains at least one internal interaction (yellow dotted lines, Fig. 2(b)
), while the other two are open (elementary structures not containing any
internal interaction).
In Fig. 3 we
display the results of the method applied to a 48mer. In this case, one
of the two local elementary substructures looks like a piece of $\beta$--sheet,
a result which testifies to the fact that within the framework of lattice
model calculations, local elementary structures can be viewed as scars of
(incipient) wild type secondary structures.

We have found that the method discussed above to read the 3D--structure of
a notional protein from its 1D-- amino acid sequence works not only for the
designed sequences which fold fast, but also in the case where the design
produces a non--folder. In
fact, in such cases, the 1D $\rightarrow$ 3D strategy (correctly) 
does not lead to a native structure.   

\newpage
\Large
Methods
\normalsize
\bigskip

{\noindent\bf 1. Selection of the sequence}

Sequences are selected within the framework of a 20 letter lattice model of
proteins, making use of Monte Carlo simulations to minimize the energy of
the chain in the native conformation with respect to the amino acid sequence
and for fixed composition \cite{sh_prl}. Operatively, the sequence is chosen among
those displaying in the native conformation a sufficiently low energy $E_n$
such that the normalized energy gap or order parameter $\xi$ 
is much larger than 1.
From these sequences, the three--dimensional native
conformation is recovered through the 3--steps algorithm described below.

{\noindent\bf 2. Search of elementary structures}

We distinguish two classes of (local) elementary structures which control,
within lattice model calculations, the phenomenon of protein folding: those
which isolated do not display any interaction (no internal interaction), and
those which do display at least one (internal) interaction (cf. Figs. 1(b),
2(b) and 3(b) ). We shall refer to them generically as local elementary
structures. When the need arises to distinguish between them, we shall call
them "open elementary structures" and "closed elementary structures",
respectively \cite{nota2}.

Candidates to the role of open elementary structures are those displaying
low values of the energy density $\epsilon_s=(j-i)^{-1}\sum_{i\leq l\leq
j}\min_{k\in\hspace{-0.15cm}|\; (i,j)} B_{\sigma(l)\sigma(k)}$, where 
$B_{\sigma(l)\sigma(k)}$ are the contact energies between the $l$th and the
$k$th amino acid of the chain (Table 6 of ref. \cite{mj}). The calculations
have been carried out with the condition $(j-i)\leq 10$. The results have
been found to be stable with respect to an increase of the range of values
of $(j-i)$.
Candidates to closed elementary structures are formed by maximizing the
value of $p(i,j)=(j-i)^{-\gamma} \exp(-\beta B_{\sigma(i)\sigma(j)})$, where
$\gamma=1.68$, $i+2\leq j \leq i+8$ and $j-i$ is odd \cite{modellino}. If local
elementary structures build more than one internal contact (cf. e.g. Fig.
3(b)), the total value of $p$ associated with them is given by the product of
the p--values associated with each of the contacts.

{\noindent\bf 3. Search of the folding core.}

The energy spectrum of the low--energy conformations which can be
constructed making use of the elementary structures 
is calculated through a complete enumeration of the conformations
having an energy smaller than a given threshold. Starting from an elementary
structure a second one is placed in all possible conformations with respect
to the first one which have at least one contact between the two, and
its energy recorded. To each
of these conformations displaying an energy lower than some chosen energy,
a new elementary structure is added, and the process
repeated until the composite system contains all elementary structures. The
calculations are repeated altering the order in which the different
elementary structures are placed together and allowed to interact with each
other. Making use of the resulting energy distribution of the conformations
containing 2,3,..., all, of the local elementary structures, we select 
as potential (post--critical) folding cores 
of the notional protein, those conformations
having an energy lying below a given threshold energy.

{\noindent\bf 4. Relaxation of the monomers not belonging to 
the core}

For each (potential) folding core, the variety of conformations of the
remaining monomers are enumerated (a number which, e.g. for the 36mer 
shown in Figs 1 and 2, is of
the order of $10^5$), and the total energy of the
system calculated. The conformation which has  energy lower than $E_c$ is
the native conformation of the notional sequence. We know that this
conformation is unique, in
keeping with the fact that the (post--critical) folding core determines, in
a unique fashion, the native conformation of a designed sequence 
\cite{aggreg,sh_nature,nucleus}.

{\noindent\bf 5. Caveats and limitations}

To calculate $E_c$ use is made of the Random Energy Model \cite{derrida}. In this
model the critical energy $E_c=N_c\sigma(2\log\gamma)^{1/2}$ depends on the
number $N_c$ of contacts of fully compact conformations (40 and 56 for chains with
$N=36$ and $N=48$, respectively), on the number of such conformations per
monomer ($\gamma=1.8$ \cite{flory}, $\gamma=2.2$ \cite{orland}) and on the variance
$\sigma$ of the contact energies
(equal to  $0.3$ for the parameters suggested in Table 6 of ref.
\cite{mj}). Using the average value $\gamma=2$ one obtains $E_c=-14.1$ and
$E_c=-19.8$ for $N=36$ and $N=48$ respectively, to be compared with the values of $-14.0$
and $-21.5$ calculated {\it a posteriori} making use of low temperature Monte Carlo simulations. 
This uncertainty on $E_c$ is of no consequence for the
workings of the method in dealing with sequences with $\xi\gg 1$, but can
limit its predictive power for sequences whose  native energy is close to
$E_c$.

\begin{figure}
\caption{Elements in the prediction of the 3D-structure of a notional protein
starting from its 1D-structure:
(a) designed 36mer sequence (1D structure) 
given as the input to solve
the 1D $\rightarrow$ 3D puzzle, together with the 20x20
contact energies among the amino acids (Table 6 of ref. \protect\cite{mj}), (b) local
elementary structures obtained following
protocol \#1 of Methods. The amino acids participating in the contacts 
of these closed structures have been drawn in (a) in white colours,
(c) only (post critical) folding core designed
making use of the elementary structures and of the
contact energies according to protocol \#3 of Methods, to which is
associated a compact conformation (shown in (d) )
obtained by relaxing the amino acids not belonging to the core (according to
protocol \#4 of Methods), displaying an
energy lower than $E_c$ ($=-14.1$, cf. protocol \#5 of Methods).
Accordingly, the conformation (d) is the predicted wild type (3D) native
conformation of the notional 1D-sequence shown
in (a) (where the hot, warm and cold sites of the protein in its native
state, calculated according to ref. [22] are displayed in terms of
red, yellow and green beads). 
In fact, this is the native conformation used in the literature to design the
sequence (a). Furthermore, 
Monte Carlo simulations testify to the correctness of the prediction.
In fact, evolving the sequence (a) starting from random elongated
conformations, it always folds into the conformation (d)
in $\approx 0.7\cdot 10^6$ MC steps. In (e) the time
evolution of all the native contacts of one particular run
(in which the chain folds in $0.65\cdot 10^6$ MC steps) is shown. In particular,
that of the contacts associated with the local elementary structures (dashed yellow
lines) and with the contacts between the
elementary structures in the folding nucleus shown in (c) (red dotted lines).
To be noted that the sequence shown in (a) and called S$_{36}$ in the
literature \protect\cite{aggreg,Tiana_98,sh_nature}, 
was designed following the inverse
folding-model according to protocol \#1 of Methods
using the structure shown in (e) as native conformation (for fixed amino
acid composition). It has an energy $E_n=-16.5$, and
thus a normalized gap $\xi = 8.3$.}
\end{figure}

\begin{figure}
\caption{Same as in Fig. 1 but 
in this case the 36mer sequence has been designed on
the native conformation (d) so as to minimize
the local contacts of the native conformation and consequently of the
folding core \protect\cite{nonlocal} (protocol \#1 of Methods). This situation which would look
particular trying for a strategy based on
local elementary structures, is solved by applying protocols \#2-5 of Methods
with the same ease than in the case of the 36mer shown in Fig. 1.
In (b) are shown the resulting local elementary structures, two of which are open.
All the amino acids participating in them are displayed in (a) in terms
of white simbols. The same colour is used for the amino acid participating
in the (internal) contacts of the closed structure.
In (c) are shown the disposition of these
structure to form the (unique) folding core with which is associated the (single)
completely compact conformation (d) with energy lower
then $E_c$. This conformation coincides with the one used in the
literature \protect\cite{nonlocal} to design the sequence shown in (a), the associated (native)
energy being $E_n=-15.99$. Furthermore, Monte Carlo simulations testify to the correctness of the
predicted native conformation (d). 
This can be seen in (e), where the time evolution of the native
contacts for a particular run is displayed as a function of the MC
steps of the folding simulations. The sequence in (a) folds into the structure shown in (d)
in approximately $2\cdot 10^6$ MC steps. The normalized gap associated with
sequence (a) is $\xi=6.3$. The hot, warm and cold sites of the protein
in its native conformation calculated according to [22] are displayed 
in terms of red, yellow and green beads respectively.}
\end{figure}

\begin{figure}
\caption{Same as in Figs. 1 and 2, but for a 48mer. With the designed sequence shown in
(a) are associated the elementary structures shown in (b), the folding core (c) with which
the native conformation shown in (d) is predicted (protocols \#2-5 of Methods). This in fact
is the conformation used in the literature \protect\cite{sh_nature} to design sequence (a)
(protocol \#1 of Methods). The correctness of the prediction is furthermore confirmed by
Monte Carlo simulations. In fact, the sequence (a) folds into the native conformation (d) in
approximately 
$3.3\cdot 10^6$ MC steps (cf. Fig. (e)). The sequence (a) is 
associated with a normalized gap $\xi=16.5$
in the native conformation (d). The amino acids shown in white in (a) are those
participating in the contacts displayed by the closed structures shown in (b).
The hot, warm and cold sites of the protein in the native conformation
calculated according to [22] are displayed in terms of red, yellow and green beads
respectively.}
\end{figure}


\begin{thebibliography}{99}
\bibitem{note1} An equivalent version of the model \protect\cite{mirny} posits that
good--folder sequences can be obtained by minimizing the time the
heteropolymer employs to target onto the chosen native conformation with
respect to amino acid sequence.
\end{thebibliography}
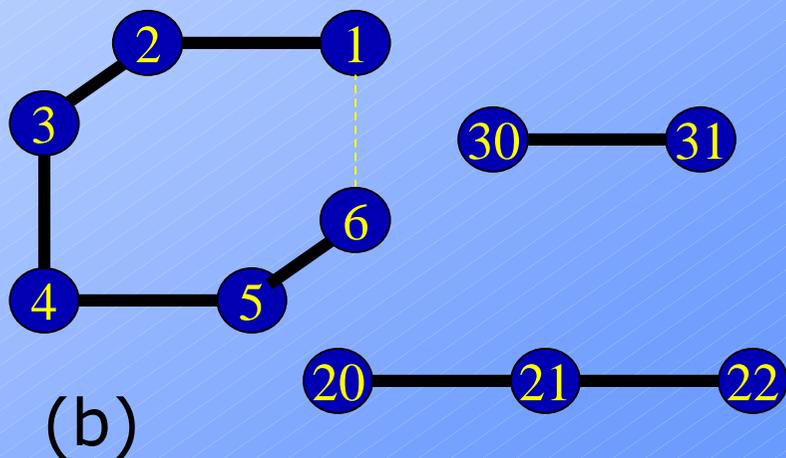
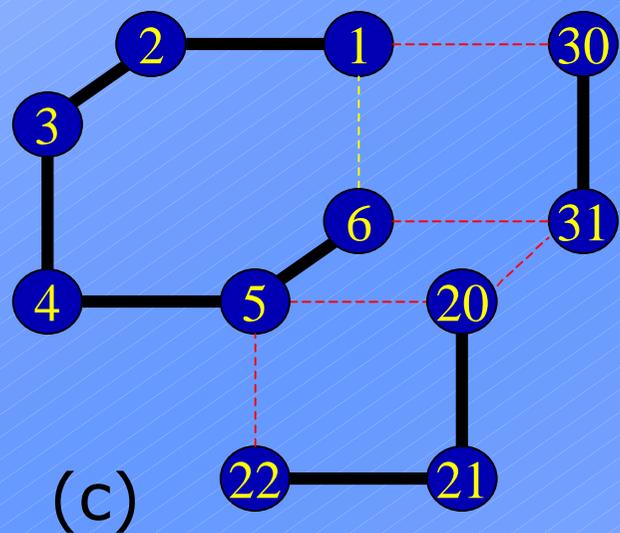
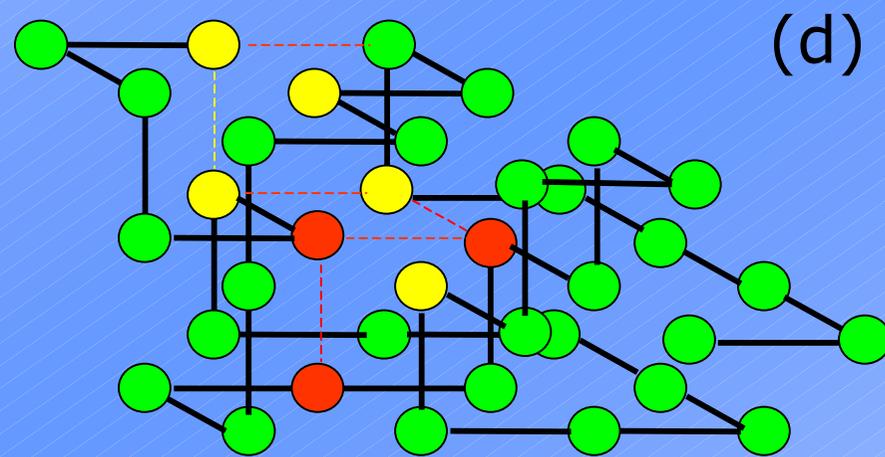
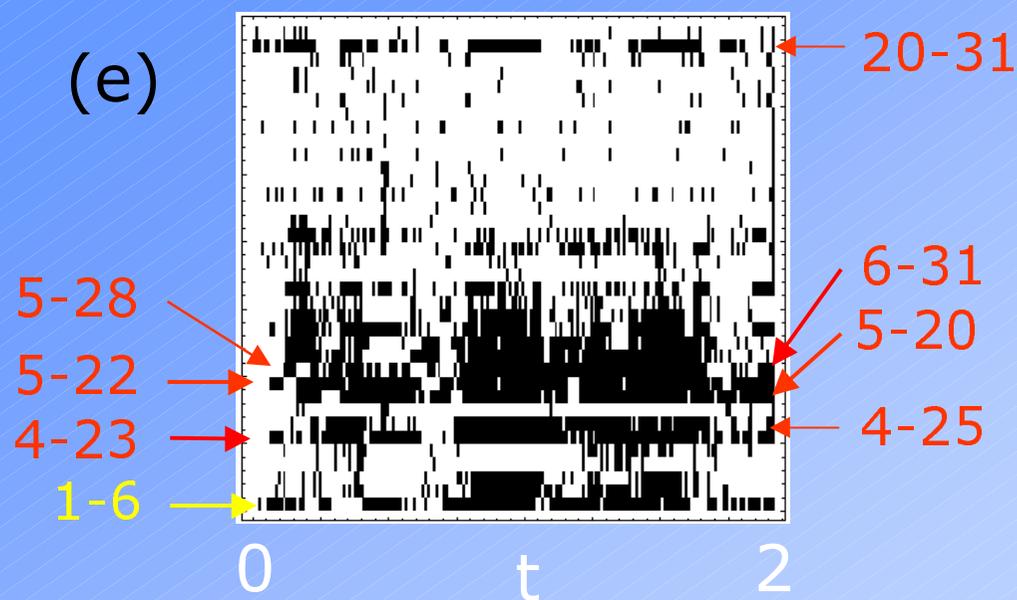
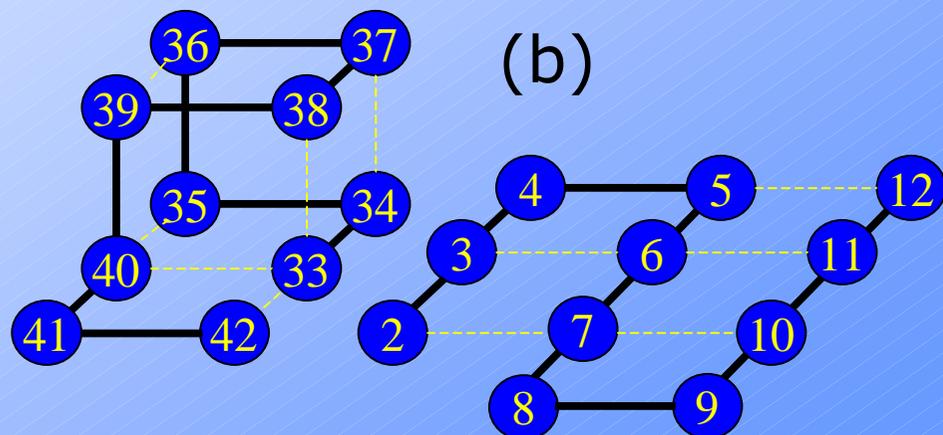
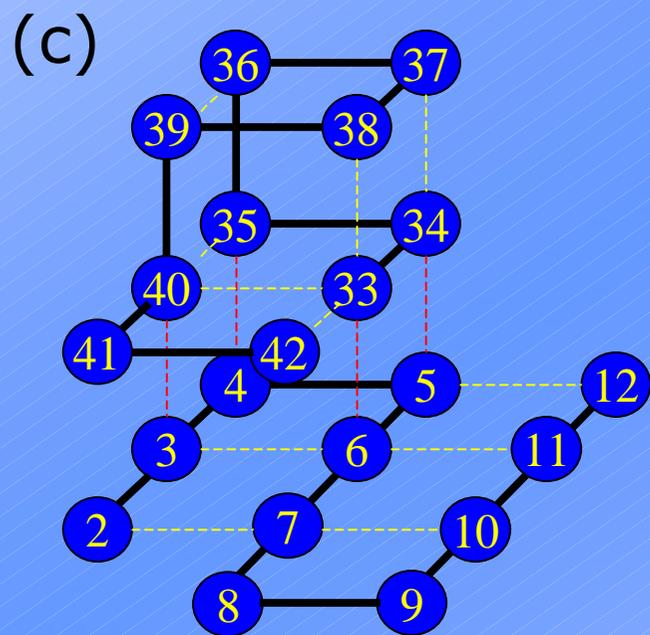
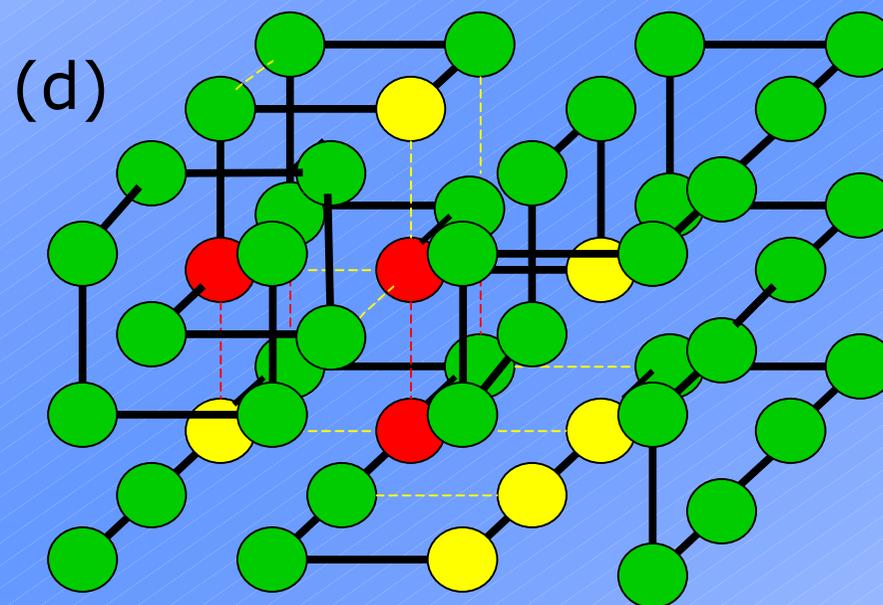
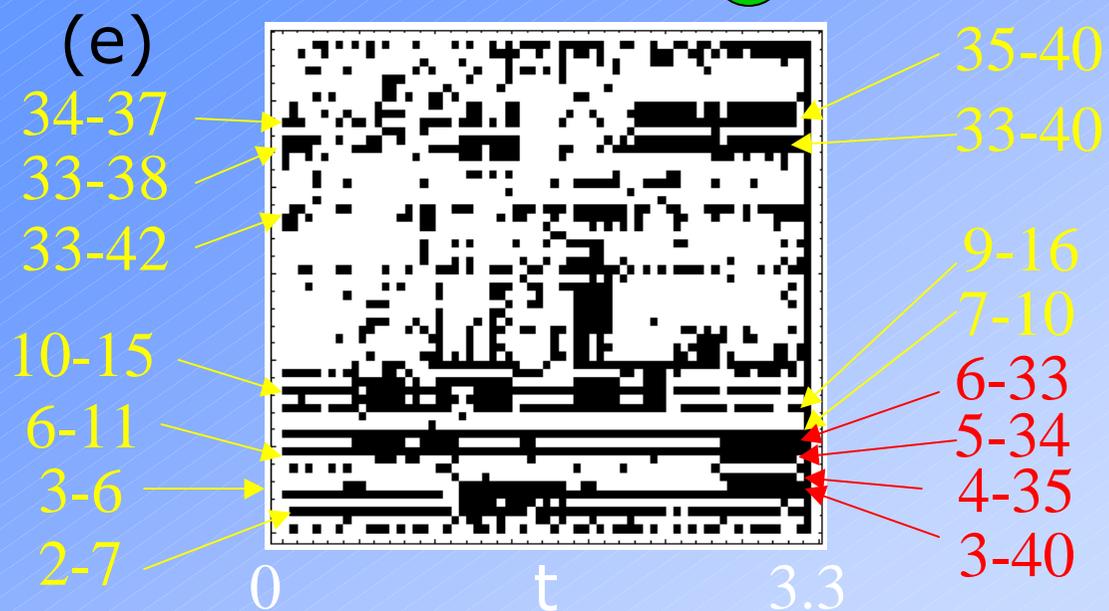
\end{document}